\documentclass[]{aa}
\usepackage{natbib}
\bibpunct{(}{)}{;}{a}{}{,} % to follow the A&A style
\usepackage{graphicx}
\usepackage{txfonts}
\usepackage{color}

\begin{document}

   \title{Dust arcs in the region of Jupiter's Trojan asteroids}

   %\subtitle{I. Overviewing the $\kappa$-mechanism}

   \author{Xiaodong Liu 
          \and
          J\"urgen Schmidt
          }

   \institute{Astronomy Research Unit, University of Oulu, Finland\\
              \email{xiaodong.liu@oulu.fi} 
             }

%   \date{}

% \abstract{}{}{}{}{} 
% 5 {} token are mandatory
 
  \abstract
   {}
   {The surfaces of the Trojan asteroids are steadily bombarded by interplanetary micrometeoroids, which releases ejecta of small dust particles. These particles form the faint dust arcs that are associated with asteroid clouds. Here we analyze the particle dynamics and structure of the arc in the region of the $L_4$ Trojan asteroids.}
   {We calculate the total cross section of the $L_4$ Trojan asteroids and the production rate of dust particles. The motion of the particles is perturbed by a variety of forces. We simulate the dynamical evolution of the dust particles, and explore the overall features of the Trojan dust arc.}
   {The simulations show that the arc is mainly composed of grains in the size range 4-10 microns. Compared to the $L_4$ Trojan asteroids, the dust arc is distributed more widely in the azimuthal direction, extending to a range of [30,120] degrees relative to Jupiter. The peak number density does not develop at $L_4$. There exist two peaks that are azimuthally displaced from $L_4$.}
   {}


   \maketitle

\section{Introduction}\label{section_introduction}
The study of the Trojan asteroids is a classical problem of celestial mechanics and astronomy, but it is also a hot topic of recent research and future space exploration (see review papers by \citet{emery2015complex}, \citet{robutel2010introduction}, \citet{dotto2008troianis}, \citet{jewitt2004jupiter}, and references therein). In 2021, the NASA mission \textit{Lucy} will explore the Trojan asteroids in a sequence of flybys. In this paper, we focus on the dynamical evolution of dust expelled from the Trojans. Dust carries information about the composition of its source, which gives important constraints on the origin and formation of the Trojan group of asteroids. Several papers studied the dynamics of dust in the Trojan region \citep{liou1995asteroidal,liou1995radiation,zimmer2014orbital,de2010studying}. However, the configuration of the dust distribution associated with the Trojan asteroids is still unclear to date. In this work, we answer this question through computer simulations of the long-term evolution of dust particles. The simulations are performed on a large computer cluster.

\section{Production rate of dust particles} \label{section_production_rate}
\begin{figure}
   \centering
   \includegraphics[width=6cm,angle=90]{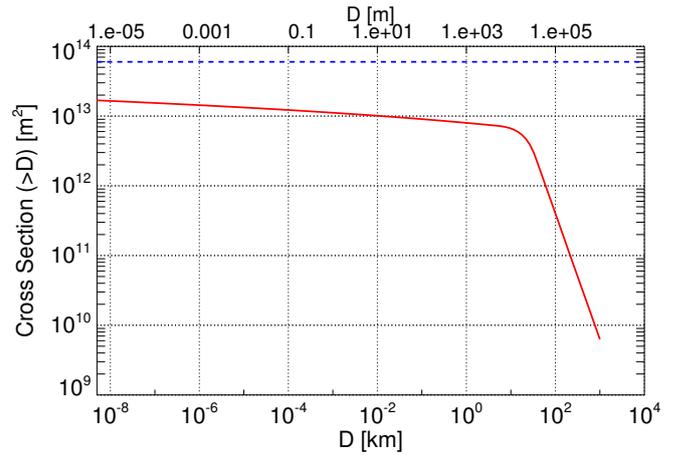}
      \caption{Total cross section of the $L_4$ Trojan asteroids (red line). The blue dashed line denotes the upper limit of $6\!\times\!10^{13} \, \mathrm{m}^2$ for the cross section of $L_5$ Trojan asteroids that are larger than $10 \, \mathrm{\mu m}$ in radius, as inferred by \citet{2000Icar..145...44K}.}         \label{fig:FernandezCross}
   \end{figure}

We use the size distribution of the $L_4$ Trojans published by \citet{2009AJ....138..240F} to estimate the total cross section of the Trojan asteroids. This is necessary to derive the production rate of dust particles. The distribution is modified from the distribution derived by \citet{Jewitt:2000ew} by taking into account the systematic dependence of the albedo on asteroid size
that has been inferred by \citet{2009AJ....138..240F}. Although this correlation was not confirmed in the much larger data set from WISE \citep{Grav:2011eq,2012ApJ...759...49G}, we continue to use the size distribution from \citet{2009AJ....138..240F} (use their Fig.~6) because it is calibrated to absolute numbers in the same way as the result by \citet{Jewitt:2000ew} and because the potential bias, if any, is small. The total asteroid cross section is obtained from the second moment of the distribution (Fig.~\ref{fig:FernandezCross}). This second moment has a logarithmic divergence toward small asteroid sizes, but for practical purposes, this is not a problem because the change in total cross section is very mild even when we vary the lower cutoff asteroid size from $10\,\mathrm{\mu m}$ to $100\, \mathrm{km}$. For our modeling we use a value of $10^{13} \, \mathrm{m}^2$.

To estimate the production rate of dust ejected from the surfaces of the Trojan asteroids by impacts of interplanetary micrometeoroids, we follow the procedure by \citet{2003P&SS...51..251K}. By
normalizing the differential distribution of ejecta particle masses to the total mass production rate $M^+$, we obtain
\begin{equation} \label{eq:pm0}
p(m)=M^+ \frac{1-\alpha}{m_\mathrm{max}^{1-\alpha}-m_\mathrm{min}^{1-\alpha}}m^{-(1+\alpha)} \,,
\end{equation}
where $m_\mathrm{max}$ and $m_\mathrm{min}$ denote the highest and lowest mass of an ejected particle. The exponent $\alpha$ is the slope of the cumulative production rate with a plausible range $0.5 \! < \! \alpha \! < \! 1$. Here we use the value 
\begin{equation}
\alpha=0.91
,\end{equation}
which was inferred from measurements on the Moon \citep{Horanyi:2015faa} of the Lunar Dust Experiment dust detector (LDEX) on board the Lunar Atmosphere and Dust Environment Explorer (LADEE) mission.

%\begin{eqnarray}
%               N^+(>m)&\equiv&\int_{m}^{m_\mathrm{max}} \mathrm{d}m\,p(m)\\
%                &=& \frac{1-\alpha}{\alpha}\frac{m^{-\alpha}-m_\mathrm{max}^{-\alpha}}{m_\mathrm{max}^{1-\alpha}-m_\mathrm{min}^{1-\alpha}} \,.
%\end{eqnarray}
It is practical to write Eq.~\ref{eq:pm0} in the form
\begin{equation} \label{eq:pm1}
p(m)=\frac{1}{m_\mathrm{max}}\,\frac{M^+}{m_\mathrm{max}}\, \frac{1-\alpha}{1-\left(\frac{m_\mathrm{min}}{m_\mathrm{max}}\right)^{1-\alpha}} \left(\frac{m}{m_\mathrm{max}}\right)^{-(1+\alpha)} \,.
\end{equation}
We frequently use particle radii $r_\mathrm{g}$ instead of mass $m$. From $p(m)\,\mathrm{d}m=p(r_\mathrm{g})\,\mathrm{d}r_\mathrm{g}$, we obtain 
\begin{equation} \label{eq:pr}
p(r_\mathrm{g}) = \frac{3}{r_\mathrm{max}} \frac{M^+}{m_\mathrm{max}} \frac{1-\alpha}{1-\left(\frac{m_\mathrm{min}}{m_\mathrm{max}}\right)^{1-\alpha}} \left( \frac{r_\mathrm{g}}{r_\mathrm{max}} \right)^{-1-3\alpha} \,.
\end{equation}
Here \begin{equation}
r_\mathrm{min}=\left(\frac{3}{4\pi \rho_\mathrm{g}} m_\mathrm{min}\right)^{1/3},\quad r_\mathrm{max}=\left(\frac{3}{4\pi \rho_\mathrm{g}} m_\mathrm{max}\right)^{1/3} \,,
\end{equation}
where $\rho_\mathrm g$ is the grain density. We note that $p(m)$ and $p(r_\mathrm{g})$ are only defined for ejecta masses $m$ with $m_\mathrm{min} \! < \! m \! < \! m_\mathrm{max}$.

For the cumulative production rate, we obtain from Eq.~\ref{eq:pr}
\begin{equation} \label{eq:prcumu}
p(> \! r_\mathrm{min}) = \int_{r_\mathrm{min}}^{r_\mathrm{max}}\mathrm{d} r_\mathrm{g} p(r_\mathrm{g})= \frac{1-\alpha}{\alpha} \, \frac{M^+}{m_\mathrm{max}}\, \frac{\left(\frac{r_\mathrm{min}}{r_\mathrm{max}}\right)^{-3\alpha}-1}{1-\left(\frac{r_\mathrm{min}}{r_\mathrm{max}}\right)^{3(1-\alpha)}}\,.
\end{equation}

To proceed, we need to quantify the total mass production rate $M^+$. We again follow \citet{2003P&SS...51..251K} and express it as
\begin{equation} \label{eq:mplus}
M^+=Y\,F_\mathrm{imp}\, S\,,
\end{equation}
where $Y$ is the yield, and $S$ is the total cross section of the target, that is, of all $L_4$ Trojans (Fig.~\ref{fig:FernandezCross}). For the projectile mass flux $F_\mathrm{imp}$, we use the value
\begin{equation} \label{eq:fimp}
F_\mathrm{imp}=10^{-15} \ \mathrm{kg} \, \mathrm{m}^{-2} \mathrm{s}^{-1} \,,
\end{equation}
which was derived by \citet{2003P&SS...51..251K} from the Divine model \citep{1993JGR....9817029D} of the interplanetary meteoroid population for a spherical target with a unit surface that is on a heliocentric circular orbit at the distance of Jupiter. For the Trojans, the effect of the gravitational focusing by a planet is not included. Newer models have been published \citep{2016Icar..264..369P}, but we continued to use the number for the mass flux from the Divine model because this has led to a quantitative match with the measured ejecta clouds for the Galilean moons \citep{Kruger:2003ik}, which are located\ at the same heliocentric distance as the Trojan asteroids.

\begin{figure}
   \centering
   \includegraphics[width=6cm,angle=90]{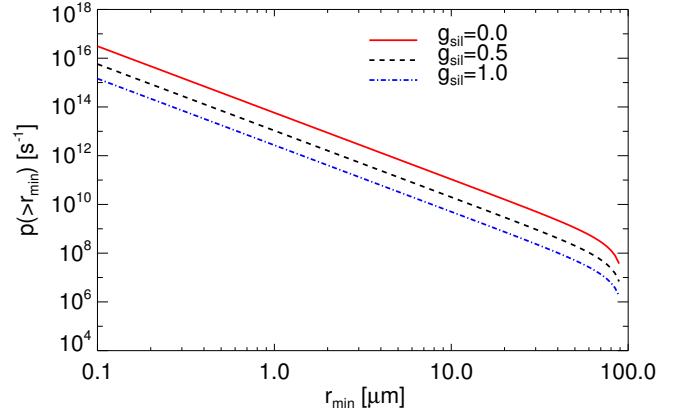}
   \caption{Cumulative production rate of ejecta particles from Eq.~\ref{eq:prcumu}. The parameter $g_\mathrm{sil}$ is the mixing ratio of silicate to ice of the target surface. The red, black, and blue lines denote the cumulative production rates for $g_\mathrm{sil}=0.0$, $g_\mathrm{sil}=0.5,$ and $g_\mathrm{sil}=1.0$, respectively.}
          \label{fig:Nplus}
   \end{figure}

For a given impact velocity $v_\mathrm{imp}$ and projectile mass $m_\mathrm{imp}$, the yield can be estimated from the empirical formula derived from laboratory experiments by \citet{2001Icar..154..391K,2001Icar..154..402K},
\begin{equation} \label{eq:yield}
\begin{split}
Y=&2.85\times10^{-8} \times 0.0149^{g_\mathrm{sil}} \times \left(\frac{1-g_\mathrm{sil}}{927}+\frac{g_\mathrm{sil}}{2900}\right)^{-1}\,\\ & \times \left(\frac{m_\mathrm{imp}}{\mathrm{kg}}\right)^{0.23} \times \left(\frac{v_\mathrm{imp}}{\mathrm{m\,s^{-1}}}\right)^{2.46}\,,
\end{split}
\end{equation}
where $g_\mathrm{sil}$ is the mixing ratio of silicate to ice of the target surface. Equation~\ref{eq:yield} assumes a density of $927 \, \mathrm{kg}/\mathrm{m}^3$ for ice and a density of $2,900 \, \mathrm{kg}/\mathrm{m}^3$ for silicate (the original formula by \citet{2001Icar..154..391K,2001Icar..154..402K} used a density of $2,800 \, \mathrm{kg}/\mathrm{m}^3$ for silicate). For $g_\mathrm{sil}=1$ we have a pure silicate surface and for $g_\mathrm{sil}=0$ we have pure ice. For our modeling we use
\begin{equation}
g_\mathrm{sil}=1
\end{equation}
because the asteroid surfaces are most probably silicate rich (even when the asteroid has a substantial interior ice fraction) and because $g_\mathrm{sil}=1$ leads to the smallest yields, which gives a lower bound for the particle number densities inferred from our model (Fig.~\ref{fig:Nplus}).

For the typical projectile mass we use \citep{2003P&SS...51..251K}
\begin{equation}
m_\mathrm{imp}=10^{-8}\ \mathrm{kg}
\end{equation}
because this particle mass (corresponding roughly to a radius of $100\,\mathrm{\mu m}$) dominates the mass flux at the distance
of Jupiter. The largest ejecta have a mass on the order of the projectile. Thus, we use 
\begin{equation}
m_\mathrm{max}=m_\mathrm{imp}=10^{-8}\ \mathrm{kg}\,.
\end{equation}
The precise choice of $m_\mathrm{max}$ has no strong effects on our results because Eq.~\ref{eq:pm0} depends only weakly on $m_\mathrm{max}$.

The average impact velocity
\begin{equation}
v_\mathrm{imp}=9\ \mathrm{km \, s^{-1}}
\end{equation}
was also calculated by \citet{2003P&SS...51..251K} from the Divine model evaluated at the distance of Jupiter.

\section{Dynamical model} \label{chap:DynamicalModel}
The motion of dust particles from the Jupiter Trojans is influenced by a variety of forces. The most important forces are solar gravity, solar radiation pressure, Poynting-Robertson drag, solar wind drag, the Lorentz force exerted by the interplanetary magnetic field, and gravitational perturbations from Jupiter and other planets (Venus, Earth, Mars, Saturn, Uranus, and Neptune). Our code integrates the equations of motion in the Jupiter orbital inertial frame $Oxyz$. Here, the $z$-axis is defined as the normal to the orbital plane of Jupiter at the J2000 epoch, and the $x$-axis is aligned with the intersection of the orbital plane and the equatorial plane of Jupiter at the J2000 epoch. The $y$-axis completes an orthogonal right-handed frame.

The equations of motion of a dust particle in the region of the Trojan asteroids read
\begin{equation} \label{equ_dynamic_model}
\begin{split}
\ddot{\vec r} = &-\frac{GM_\mathrm{S}}{r^3}{\vec r} + \sum_{i=1}^{7}GM_{\mathrm P_i}\left(\frac{\vec r_{\mathrm {dP}_i}}{r_{\mathrm {dP}_i}^3}-\frac{\vec r_{\mathrm P_i}}{r_{\mathrm P_i}^3}\right) + \frac{Q}{m}\left(\dot{\vec r}-\vec v_{sw} \right)\times{\vec B} \\
& + \frac{3Q_\mathrm SQ_\mathrm {pr}\mathrm{AU}^2}{4r^2\rho_\mathrm gr_\mathrm gc}\left\{\left[1-(1+sw)\frac{\dot r}{c}\right]\hat{\vec r} - (1+sw)\frac{\dot{\vec r}}{c}\right\} \,.
\end{split}
\end{equation}
Here, ${\vec r}$ is the heliocentric radius vector of the grain, $G$ is the gravitational constant, $M_\mathrm S$ is the mass of the Sun, $M_{\mathrm P_i}$ the mass of the $i$th planet, $\vec r_{\mathrm {dP}_i}$ the vector from the particle to the $i$th planet, $\vec r_{\mathrm P_i}$ the vector from the Sun to the $i$th planet, $Q=4\pi\varepsilon_0r_\mathrm g\Phi$ is the grain charge with the vacuum permittivity $\varepsilon_0$, $\Phi$ is
the grain surface potential, $\vec v_{sw}$ is the solar wind velocity, $\vec B$ is the interplanetary magnetic field, $Q_\mathrm S$ is the solar radiation energy flux at one AU (astronomical unit), $Q_\mathrm{pr}$ is the solar radiation pressure efficiency factor, $c$ is the speed of light, and $sw$ is the ratio of solar wind drag to the Poynting-Robertson drag, which depends on $Q_\mathrm{pr}$ \citep{gustafson1994physics}.

\begin{figure}
    \centering 
    \includegraphics[width=6cm,angle=90]{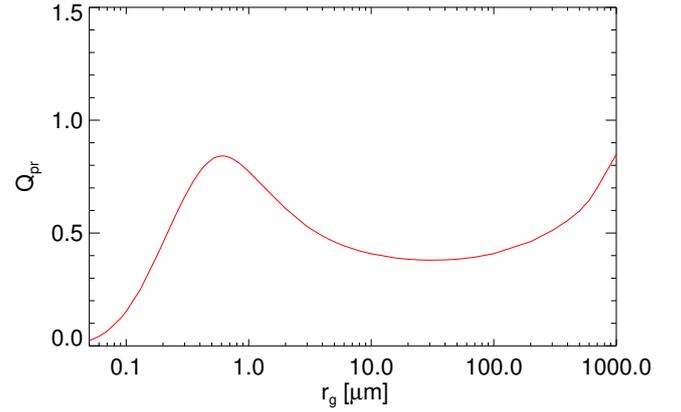}    \caption{Size-dependent radiation pressure efficiency $Q_\mathrm{pr}$. The values are calculated based on the Mie theory \citep{mishchenko1999bidirectional, mishchenko2002scattering}, using optical constants for silicate \citep{mukai1989cometary}.}
    \label{fig_MakeQprMukai}
    \end{figure}

The parameterization of the interplanetary magnetic field described in \citet{gustafson1994physics} and \citet{landgraf2000modeling} is used, and a constant surface potential +5 V \citep{1996Sci...274.1501Z} is adopted. To calculate $Q_\mathrm{pr}$, the optical constants for silicate grains are taken from \citet{mukai1989cometary}. The dependence of $Q_\mathrm{pr}$ on grain size for silicate particles, calculated based on the Mie theory \citep{mishchenko1999bidirectional, mishchenko2002scattering} for spherical grains, is shown in Fig.~\ref{fig_MakeQprMukai} (also compare to Fig.~8 of \citet{krivov2002dust}).

As a sink for dust particles, impacts on the planets are considered. The probability of a close encounter of a particle with a Trojan asteroid is extremely low. The total cross section of $L_4$ Trojan asteroids is about $10^{13} \, \mathrm{m}^2$ (Fig.~\ref{fig:FernandezCross}). With a plausible extension of the $L_4$ cloud of $(40^\circ \! \times \! 5\,\mathrm{AU})\times(1\,\mathrm{AU})$ in the Jovian orbital plane, this means an optical depth of about $\tau=10^{-10}$, which translates into a characteristic collision time for dust with a Trojan asteroid that is longer than the lifetime of the solar system. Thus, the Trojan asteroids can be safely neglected as sinks and as gravitational perturbers. We also stop the integration when the distance between the particle and the $z$-axis is smaller than $0.5 \, \mathrm{AU}$ or larger than $15 \, \mathrm{AU}$, in which case we assume that the particle has escaped from the region of interest. This escape efficiently acts as another sink in our model.

\section{Simulational scheme}
Integrations for ten different grain sizes (radii) are carried out: $0.5 \,\mathrm{\mu m}$, $1 \, \mathrm{\mu m}$, $2 \, \mathrm{\mu m}$, $4 \, \mathrm{\mu m}$, $5 \, \mathrm{\mu m}$, $6 \, \mathrm{\mu m}$, $8 \, \mathrm{\mu m}$, $10 \, \mathrm{\mu m}$, $16 \, \mathrm{\mu m,}$ and $32 \, \mathrm{\mu m}$. For each grain size, 100 particles are started from randomly selected 100 $L_4$ Trojan asteroids and integrated forward in time. The initial orbits of the particles are assumed to be the same as their respective source asteroids. The osculating orbital elements of these Jupiter Trojans at the launching time of the particles are obtained from the JPL Small-Body Database Search Engine. The motions of particles are followed until each of them hit a sink. To handle the heavy computational load, these long-term simulations are performed on the large computer cluster located at the Finnish CSC--IT Center for Science.

We do not store the rapidly changing phase-space coordinates of particles during their evolution because of the huge storage space required. Instead, the slowly changing osculating orbital elements are stored, including the semi-major axis $a$, the
eccentricity $e$, the inclination $i$, the argument of pericenter $\omega$, and the longitude of ascending node $\Omega$. Generally, we store the set of elements 100 times per orbit. The storage space required in total is about 280 GB. We assume that the orbital segment between two consecutive stored times $t_j$ and $t_{j+1}$ is Keplerian (with constant values of $a$, $e$, $i$, $\omega,$ and $\Omega$, but different values of the true anomaly $f$). The segment is further divided into $m$ segments of length $\Delta t$. The time interval $\Delta t$ is constant for all trajectories in the whole simulation to ensure that particle positions are stored equidistantly in time. Each storage corresponds to a particle in the simulation \citep{liu2016dynamics}.

The grain trajectories are transformed from the inertial frame $Oxyz$ into a rotating frame $Ox_\mathrm{rot}y_\mathrm{rot}z$ in order to evaluate the spatial configuration of dust. The frame $Ox_\mathrm{rot}y_\mathrm{rot}z$ shares the $z$-axis with $Oxyz$. The $x_\mathrm{rot}$-axis always points from the Sun to Jupiter, and the $y_\mathrm{rot}$-axis completes an orthogonal right-handed frame. Cylindrical coordinates $(\rho$, $\phi_\mathrm{rot}$, $z)$ in the rotating frame $Ox_\mathrm{rot}y_\mathrm{rot}z$ are defined such that $\rho$ = $\sqrt{x_\mathrm{rot}^2+y_\mathrm{rot}^2}$ and $\phi_\mathrm{rot}$ = $\mathrm{atan2}(y_\mathrm{rot},x_\mathrm{rot})$.

The Trojan region is divided into a number of cylindrical grid cells. For each particle, we determine the cell index $(i_\mathrm{cell}, j_\mathrm{cell}, k_\mathrm{cell})$ where the particle is located. The phase-space number density reads
\begin{equation} \label{equ_cumu_density}
n(i_\mathrm{cell}, j_\mathrm{cell}, k_\mathrm{cell}; > \! 0.5 \,\mathrm{\mu m}) = \int_{0.5 \,\mathrm{\mu m}}^{32 \, \mathrm{\mu m}} \mathrm{d}r_\mathrm{g} \, p(r_\mathrm{g}) \Delta t \, \frac{\tilde n (i_\mathrm{cell}, j_\mathrm{cell}, k_\mathrm{cell}; r_\mathrm{g})}{n_\mathrm{start}(r_\mathrm{g})} \,.
\end{equation}
Here $\mathrm{d}r_\mathrm{g} \, p(r_\mathrm{g})$ is the number of particles of size in the range $[r_\mathrm{g}, \, r_\mathrm{g}\!+\!\mathrm{d}r_\mathrm{g}]$ that are produced per second in the region of the $L_4$ Trojans, $p(r_\mathrm{g})$ is defined by Eq.~\ref{eq:pr}, and $\tilde n (i_\mathrm{cell}, j_\mathrm{cell}, k_\mathrm{cell}; r_\mathrm{g})$ is the number of particles with grain size $r_\mathrm{g}$ in the cell $(i_\mathrm{cell}, j_\mathrm{cell}, k_\mathrm{cell})$ divided by the cell volume, and $n_\mathrm{start}(r_\mathrm{g})=100$ is the number of particles started for each size.

\section{Numerical results} \label{subsec_numerical_results}
\begin{figure}
\centering 
\includegraphics[width=9cm]{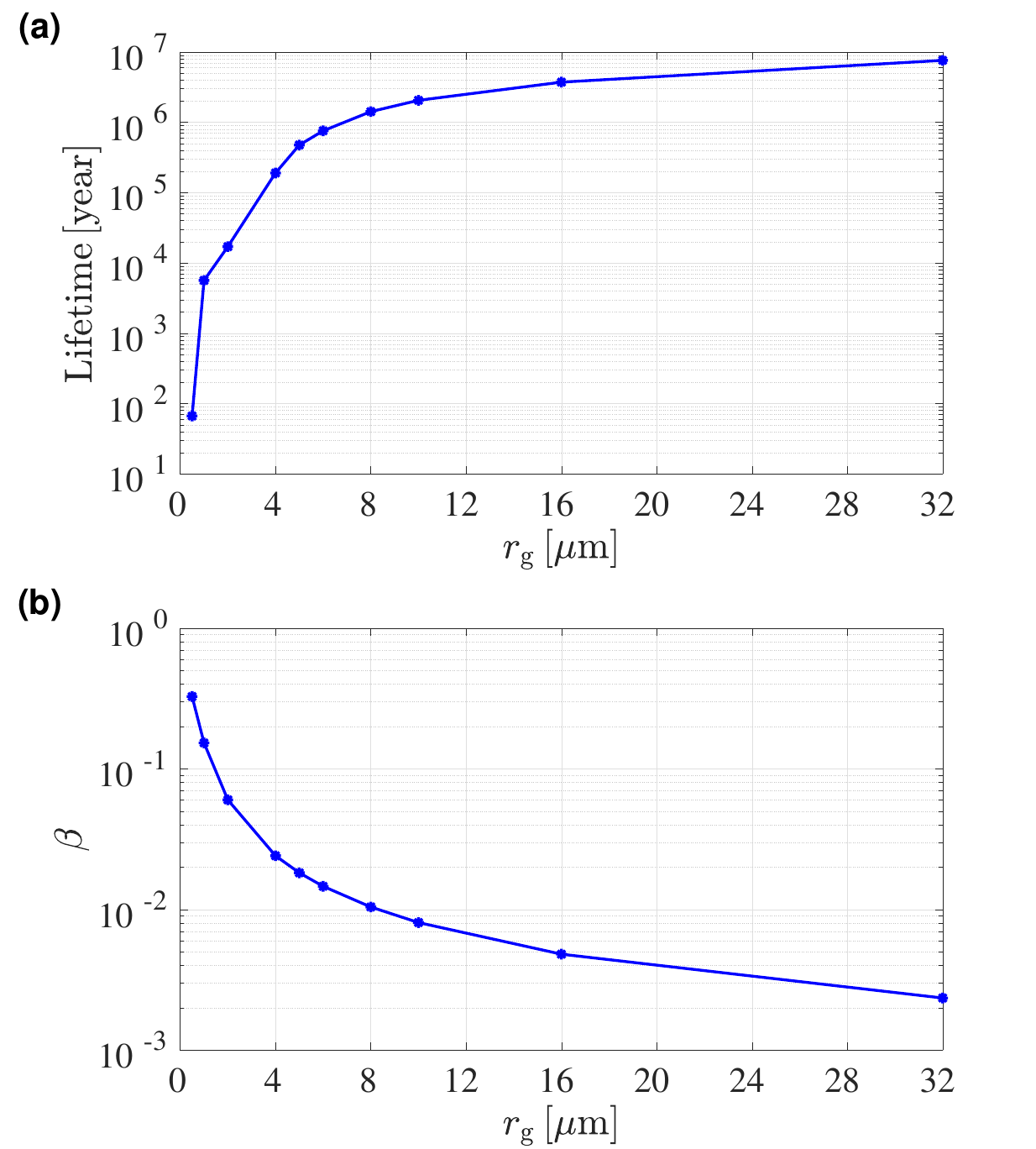}
\caption{\textbf{(a)} Average lifetimes of simulated particles as a function of grain size. \textbf{(b)} The values of $\beta$ as a function of grain size.}
\label{fig:Lifetime}
\end{figure}

The average lifetimes as a function of grain size can be directly determined from the numerical simulations (Fig.~\ref{fig:Lifetime}(a)). This can be understood in terms of the parameter $\beta$, which is defined as the ratio of solar radiation pressure and solar gravitation \citep{BURNS:1979wg},
\begin{equation} \label{equ_beta}
\beta = \frac{3Q_\mathrm SQ_\mathrm {pr}\mathrm{AU}^2}{4GM_\mathrm{S}\rho_\mathrm gr_\mathrm gc} \,.
\end{equation}
Figure~\ref{fig:Lifetime}(b) shows the values of $\beta$ for different grain sizes. For particles in the size range $[0.5, 2] \, \mu\mathrm{m}$, the value of $\beta$ is high, implying a strong perturbation by solar radiation pressure, which induces high values of the effective semi-major axis and eccentricity \citep{liou1995asteroidal}. As a result, these small particles have short lifetimes from tens of years to tens of thousands
of years. With increasing grain size, the value of $\beta$ becomes lower, and therefore the gravity of the Sun (and Jupiter) becomes dominant. These particles have longer lifetimes, from one hundred thousand years to several millions of years. Most of the particles in the Trojan region are finally transported outward to a distance $> \! 15 \, \mathrm{AU}$. A few of them are transported inward to a distance $< \! 0.5 \, \mathrm{AU}$. Very few particles hit Jupiter.

\begin{figure}
\centering 
\includegraphics[width=6cm,angle=90]{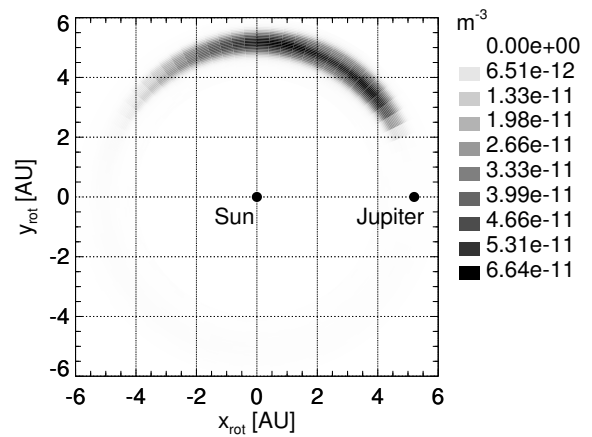}
\caption{Dust number density from the simulations of grains from the $L_4$ Trojan asteroids in the rotating frame $Ox_\mathrm{rot}y_\mathrm{rot}z$. The cumulative number density is calculated for $r_\mathrm{g} \!> \!0.5 \, \mathrm{\mu m}$, vertically averaged over $[-0.55, 0.55] \, \mathrm{AU}$. The Sun in is located at the origin and Jupiter is close to $5.2 \, \mathrm{AU}$ on the $x_\mathrm{rot}$-axis.}
\label{fig:SimuPolar}
\end{figure}

The number density based on Eq.~\ref{equ_cumu_density} for the size range $[0.5, 32] \, \mu\mathrm{m}$ in the $x_\mathrm{rot}-y_\mathrm{rot}$ plane is shown in Fig.~\ref{fig:SimuPolar}, vertically averaged over $[-0.55, 0.55] \, \mathrm{AU}$. Jupiter lies on the $x_\mathrm{rot}$-axis, that is, at $\phi_\mathrm{rot}=0^\circ$. The arc spans a wide azimuthal range $30^\circ \! < \! \phi_\mathrm{rot} \! < \! 120^\circ$, much wider than the range of the Trojan asteroids. The reason
for this is mainly solar radiation pressure and the drag forces experienced by the dust particles.

\begin{figure}
\centering 
\includegraphics[width=9cm]{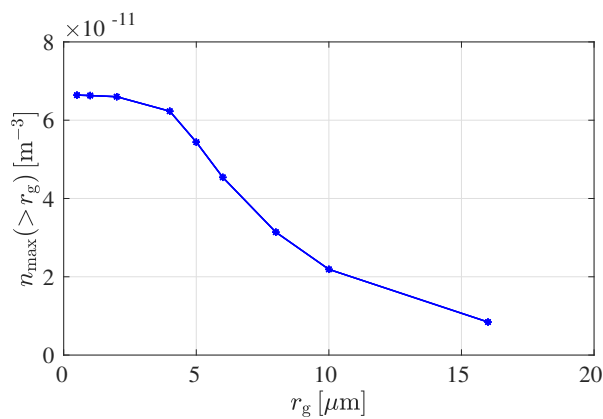}
\caption{Peak values of cumulative dust number density from Fig.~\ref{fig:SimuPolar} for different minimum particle sizes $r_\mathrm{g}$.}
\label{fig:SimuSizeDist}
\end{figure}

The peak number densities from Fig.~\ref{fig:SimuPolar} for different particle sizes are shown in Fig.~\ref{fig:SimuSizeDist}. In this cumulative plot, a steeper gradient corresponds to a larger contribution to the number density. Thus, the particles in the size range $[4,10]\, \mu\mathrm{m}$ contribute most to the number density, that is, they are most common in the Trojan arc. These particles stay in the Trojan region for about $10^5\!-\!10^6$ years (Fig.~\ref{fig:Lifetime}(a)). For particles with grain size $r_\mathrm{g} \! < \! 2 \, \mu \mathrm{m}$, the curve is almost flat, which implies that the contribution of particles with grain size $r_\mathrm{g} \! < \! 2 \, \mu \mathrm{m}$ to the number density is very small. The reason is that the strong solar radiation pressure (Fig.~\ref{fig:Lifetime}(b)) rapidly expels the small particles from the Trojan region.

\begin{figure}
\centering 
\includegraphics[width=9cm]{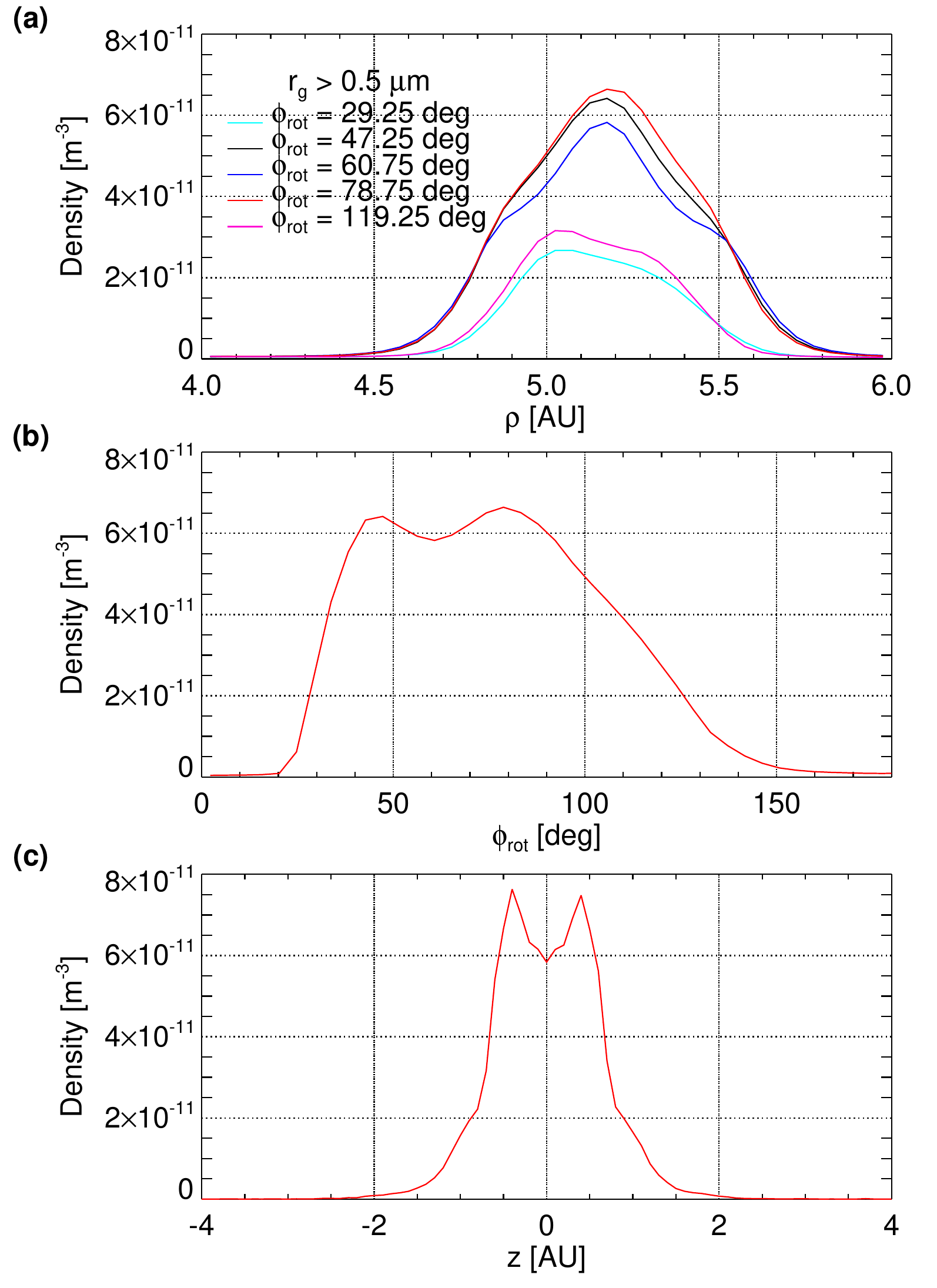}
\caption{\textbf{(a)} Radial profiles of the dust number density at different azimuthal angles from the position of Jupiter. \textbf{(b)} Azimuthal profile of the dust number density for $\rho = 5.175 \, \mathrm{AU}$ from Fig.~\ref{fig:SimuPolar}. \textbf{(c)} Vertical cut through the densest part ($\rho = 5.175 \, \mathrm{AU}$ and $\phi_\mathrm{rot} = 78.75^{\circ}$) of the dust configuration.}
\label{fig:SimuCut}
\end{figure}

The radial profiles of dust number density at different longitudes are shown in Fig.~\ref{fig:SimuCut}(a). Interestingly, the peak is not at $L_4$ ($\phi_\mathrm{rot}=60^\circ$), but close to $\phi_\mathrm{rot}=78.75^\circ$ and $\phi_\mathrm{rot}=47.25^\circ$, with a projected distance of about $\rho=5.175 \, \mathrm{AU}$. For the radial profiles along $\phi_\mathrm{rot}=29.25^\circ$ and $\phi_\mathrm{rot}=119.25^\circ$, the peak locations are close to $\rho=5.025 \, \mathrm{AU}$. The azimuthal profile of the dust number density at $\rho=5.175 \, \mathrm{AU}$ is shown in Fig.~\ref{fig:SimuCut}(b). There is a local minimum around the $L_4$ point ($\phi_\mathrm{rot}=60^\circ$). Based on Fig.~\ref{fig:SimuCut}(b), we confirm that the maxima are located close to $\phi_\mathrm{rot}=78.75^\circ$ and $\phi_\mathrm{rot}=47.25^\circ$.

A vertical cut through the densest part of the dust configuration ($\rho=5.175 \, \mathrm{AU}$, $\phi_\mathrm{rot}=78.75^\circ$) is shown in Fig.~\ref{fig:SimuCut}(c). The dust configuration is widely spread out in the vertical direction in the range $[-1, 1] \, \mathrm{AU}$, with peaks at about $z=\pm 0.4 \, \mathrm{AU}$. The peak values shown in Figs.~\ref{fig:SimuPolar} and \ref{fig:SimuCut}(a,b) are slightly lower than those in Fig.~\ref{fig:SimuCut}(c) because Figs.~\ref{fig:SimuPolar} and \ref{fig:SimuCut}(a,b) show a number density that is vertically averaged over $[-0.55, 0.55] \, \mathrm{AU}$. There is also a local minimum near the mid-plane in Fig.~\ref{fig:SimuCut}(c).

\section{Comparison with observations and prospect for detection}
An upper limit for the dust number density in the Trojan region can be obtained from the upper limit on the infrared flux inferred from the Cosmic Background Explorer (COBE) satellite data reported by \citet{2000Icar..145...44K}. The authors estimated that the cross section of material in the region of $L_5$ is no more than $6 \! \times \! 10^{13} \, \mathrm{m}^2$. Because the COBE measurement was made at $60 \,\mu\mathrm{m}$ and only bodies with radii fulfilling $2\pi r_\mathrm{g} \! \gtrsim \! \lambda$ contribute to the flux, the measurements constrain the cross section of particles larger than roughly $10\, \mu\mathrm{m}$ in radius \citep{Jewitt:2000ew}. Thus, the COBE non-detection of brightness in the $L_5$ region implies that the number of $10 \, \mu\mathrm{m}$ particles is smaller than $2 \! \times \! 10^{23}$. We use a volume for the Trojan region (at either $L_4$ or $L_5$) of roughly ($40^\circ \! \times \! 5\,\mathrm{AU})\times(1\,\mathrm{AU})\times(0.2\,\mathrm{AU})\approx 2\times10^{33} \, \mathrm{m}^3$ (azimuthal$ \, \times \, $radial extent$\, \times \, $vertical). This gives a rough upper limit for the number density of particles with $r_\mathrm{g} \! > \! 10 \, \mu\mathrm{m}$ 
\begin{equation} \label{eq:UpperLimit}
n(> \! 10 \, \mu\mathrm{m}) \lesssim 10^{-10} \,\mathrm{m}^{-3}\,.
\end{equation}
According to our simulations (Section \ref{subsec_numerical_results}), the peak density of grains larger than $0.5\,\mu \mathrm{m}$ in radius should be around $6.6 \! \times \! 10^{-11} \, \mathrm{m}^{-3}$, and for grains larger than $10\,\mathrm{\mu m}$ in radius, it should be around $2.2 \! \times \! 10^{-11} \, \mathrm{m}^{-3}$ (Fig.~\ref{fig:SimuSizeDist}). This is fairly close to but still consistent with the upper limit of $10^{-10} \, \mathrm{m}^{-3}$ for $r_\mathrm{g} \! > \! 10\,\mathrm{\mu m}$ grains that was derived from COBE data.

The amount of dust produced in collisions of Trojans was estimated by \citet{2010A&A...512A..65D}, who used a collisional fragmentation code to follow the evolution of a Trojan swarm. Their Eq.~16 together with their Fig.~5 gives for dust in the diameter range of $5\!-\!500 \, \mu \mathrm{m}$ a cross section of $10^{18} \, \mathrm{m}^2$ for the $L_4$ Trojan clouds. This is more than four orders of magnitude larger than the upper limit of $6 \! \times \! 10^{13} \, \mathrm{m}^2$ derived by \citet{2000Icar..145...44K}. One possible explanation is that \citet{2010A&A...512A..65D} did not include the direct solar radiation pressure in their model and therefore obtained dust lifetimes that are too long.

To estimate the particle counts that are expected for a dust detector on a spacecraft, we calculate column number densities of grains larger than $0.5 \, \mathrm{\mu m}$ along characteristic paths through the dust configuration. First, for a radial path in the mid-plane at an angle of $\phi_\mathrm{rot}=78.75^\circ$ from Jupiter (corresponding to the location of the maximum dust number density in the azimuthal direction, see Figs.~\ref{fig:SimuPolar} and \ref{fig:SimuCut}(a,b)), we obtain about 7 particles per $\mathrm{m^2}$. Second, along a vertical path at $\rho=5.175 \, \mathrm{AU}$ and $\phi_\mathrm{rot}=78.75^\circ$ (passing through the location of highest dust density), we obtain about 19 particles per $\mathrm{m^2}$. These numbers are for a pure silicate surface of the Trojans. If the surfaces contain a fraction of ice, the production rate will be higher (Fig.~\ref{fig:Nplus}). For a pure ice surface, the particle counts would be higher by more than one order of magnitude than for the pure silicate case, assuming the same lifetimes for ice and silicate particles.

\section{Conclusions} \label{section_conclusions}
We have analyzed the properties of a faint dust population associated with the Jupiter $L_4$ Trojan asteroids. With massive simulations, we find that particles in the size range of [4,10] microns are dominant. We derive the overall shape of the steady-state dust configuration. Compared to their sources, that is, the Trojan asteroids, the dust particles are distributed more widely in the azimuthal direction, covering a range of $[30^{\circ},120^{\circ}]$ relative to Jupiter. The peak number density does not lie at the $L_4$ point, but close to $78.75^{\circ}$ and $47.25^{\circ}$ in the azimuthal direction relative to Jupiter. Dust particles are also widely distributed in the vertical direction, with peaks of the number density located at about $z=\pm 0.4 \, \mathrm{AU}$. There is a local minimum of the number density around the $L_4$ point and the mid-plane. We expect similar properties of the dust distribution in the region of the $L_5$ Trojan asteroids because the dynamical environment is similar.

\begin{acknowledgements}
      This work is supported by the \emph{European Space Agency} under the project Jovian Micrometeoroid Environment Model (JMEM) (contract number: 4000107249/12/NL/AF). We acknowledge the \emph{CSC -- IT Center for Science} for the allocation of computational resources on their Taito cluster. We thank Heikki Salo for helpful discussions. We thank Sascha Kempf for his review and useful comments.
\end{acknowledgements}

%\bibliographystyle{aa} % style aa.bst
%\bibliography{Strings,AdditionalLit,PapersLit,lit,Ring2Galilean,trojan} % your references Yourfile.bib 

\end{document}